# Silicon-based Intermediate-band Infrared Photodetector realized by Te Hyperdoping

*Mao Wang\*, Eric García-Hemme, Yonder Berencén, René Hübner, Yufang Xie, Lars Rebohle, Chi Xu, Harald Schneider, Manfred Helm and Shengqiang Zhou\**

Dr. M. Wang, Dr. Y. Berencén, Dr. R. Hübner, Y. Xie, Dr. L. Rebohle, Dr. C. Xu, Dr. H. Schneider, Prof. M. Helm, Dr. S. Zhou
Institute of Ion Beam Physics and Materials Research, Helmholtz-Zentrum Dresden-Rossendorf, 01328 Dresden, Germany
[\*] E-mail: m.wang@hzdr.de, s.zhou@hzdr.de

Dr. Eric García-Hemme,
Dpto. de Estructura de la Materia, Física Térmica y Electrónica, Univ. Complutense de Madrid, 28040 Madrid, Spain

Y. Xie, Prof. M. Helm
Faculty of Physics and Center for Advancing Electronics Dresden Technische Universität Dresden 01062 Dresden, Germany

Keywords: Si photonics, mid-wavelength infrared photodetector, hyperdoping, ion implantation, CMOS compatible

**Abstract**

Si-based photodetectors satisfy the criteria of low-cost and environmental-friendly, and can enable the development of on-chip complementary metal-oxide-semiconductor (CMOS)-compatible photonic systems. However, extending their room-temperature photoresponse into the mid-wavelength infrared (MWIR) regime remains challenging due to the intrinsic bandgap of Si. Here, we report on a comprehensive study of a room-temperature MWIR photodetector based on Si hyperdoped with Te. The demonstrated MWIR *p-n* photodiode exhibits a spectral photoresponse up to 5 µm and a slightly lower detector performance than the commercial devices in the wavelength range of 1.0-1.9 µm. We also investigate the correlation between the background noise and the sensitivity of the Te-hyperdoped Si photodiode, where the maximum room-temperature specific detectivity is found to be $3.2 \times 10^{12}$ cmHz$^{1/2}$W$^{-1}$ and $9.2 \times 10^{8}$ cmHz$^{1/2}$W$^{-1}$ at 1 µm and 1.55 µm, respectively. This work contributes to pave the way towards establishing a Si-based broadband infrared photonic system operating at room temperature.

# 1. Introduction

Broadband infrared photodetectors have been attracting the interest of many researchers due to their wide variety of applications, such as telecommunication, security equipment, environmental sensing, and biomedicine.[1-5] Nowadays, commercially available photodetectors are mostly fabricated with mercury cadmium telluride (MCT, HgCdTe), PbS and III-V quantum-well/dot (QWIP/QWID).[6-13] These photodetectors exhibit high device performance in infrared light detection, but suffer from some crucial drawbacks, such as high cost, problematic environmental impact, operation at cryogenic temperatures and especially incompatibility with the complementary metal-oxide-semiconductor (CMOS) fabrication routes. Alternatively, Si-based photodetectors overcome these disadvantages, [8,14] but their infrared photoresponse is fundamentally restricted to the near infrared (NIR) spectral range due to the 1.12-eV indirect bandgap of Si ($\lambda$ = 1.1 μm). Therefore, the development of a room-temperature broadband infrared Si-based photodetector is of great interest in the realm of all-Si photonic systems.

One of the most promising approaches to further extend the room-temperature optical response of Si to the short- and mid-wavelength infrared (SWIR, MWIR) range consists of introducing deep-level dopants (e.g. transition metals and chalcogen dopants) into Si at concentrations in excess of the solid solubility limit.[15-25] This process leads to the broadening of the deep-level states into an intermediate band (IB) with finite width that allows for the strong optical absorption of photons with an energy lower than that of the Si bandgap.[26] Moreover, utilizing deep-level impurities provides a path to obtain extrinsic Si-based photodetectors for room-temperature operation, which is not possible with shallow-level impurities.[15,18,19,21-23] However, deep-level impurities such as transition metals have very high diffusion velocities in Si.[27] Therefore, they will diffuse to the surface and form the so-called cellular breakdown, preventing the dopant incorporation and eventually the fabrication of CMOS-compatible devices.[28] Si hyperdoped with chalcogen dopants does not exhibit celluar breakdown and has also shown potential applications for infrared photodetectors.[15,22] Unfortunately, due to the lack of a meticulous design of the device architecture and of the doping homogeneity, the achieved S- or Se-hyperdoped Si photodiodes show quite low sub-bandgap external quantum efficiency (EQE) and only operate in the SWIR wavelength range. Moreover, S- and Se-hyperdoped Si lose their ability to absorb infrared light even after short-duration thermal treatments at low temperatures. [29,30] This will compromise the application of hyperdoped Si for integration with the existing CMOS-compatible processes involving temperature-dependent steps. Unlike S and Se, Te impurities have much lower diffusivity.[12] Te-hyperdoped Si shows stable IR-absorption upon thermal processing up to 400 °C with a duration of 10 min.[29,32] Therefore, Si hyperdoped with Te holds promise for fabricating photodetectors with broad spectral response and enhanced detectivity as well as their integration into manufacturing processes.

In this work, we report on a room-temperature MWIR Si *p-n* photodiode working in photovoltaic mode based on Si hyperdoped with Te. The hyperdoped Si layers are homogeneous, free of cellular breakdown and surface flat. The materials and device fabrication are fully CMOS-technology compatible. The fabricated photodiode exhibits an enhanced performance figure of merit, e.g. spectral photoresponse, specific detectivity, bandwidth and response speed. These results point out the potential of Te-hyperdoped intermediate-band Si photodetectors for room-temperature high-performance MWIR detection as the new generation of Si-based photonic systems.

## 2. Results and discussion
### 2.1. Material characterization

The process flow of fabricating the Te-hyperdoped Si photodetector is shown in **Figure 1**(a)-(b). Detailed microstructure investigations of the Te-hyperdoped Si layer were carried out by Rutherford backscattering spectrometry (RBS) and transmission electron microscopy (TEM).[32-34] Single-crystalline regrowth of the Te-hyperdoped Si with a flat sample surface is achieved by the pulsed laser melting (PLM) treatment (as shown in Figure 1(c)-(f)). Moreover, Te is found to be uniformly distributed within the top 120 nm of the Si wafer without the formation of extended defects, secondary phases, or cellular breakdown. Figure 1(g) shows the absorption spectrum for the PLM-treated Te-hyperdoped Si sample. A virgin Si sample is also measured as a reference. More detailed information about the calculation of the absorption coefficient ($\alpha$) can be found in the supporting information (SI). As described in our previous work,[34] the Te-hyperdoped Si layer exhibits strong broadband sub-bandgap optical absorption in the MIR region, as compared with a bare Si sample. This is consistent with the previously reported results about S- and Se-hyperdoped Si.[22,35] In particular, a well-defined broad absorptance band peaking at around 0.36 eV is observed, which is attributed to the presence of an intermediate band.

To get insight into the optical capture cross-section $\sigma$ and the probability distribution of the binding energy $E_{Te}$ of Te-induced deep-level states, the absorption spectrum is fitted by

$$\alpha_e = N_{Te}\sigma_e \quad (1)$$

where $N_{Te}$ is the concentration of Te states and $\alpha_e$ is the absorption coefficient for the sub-bandgap absorption by assuming that electrons are promoted from Te deep-level states to the conduction band (CB).[36] This assumption is reasonable since the optical transition from the IB to the CB is more intense than that of the valence band (VB) to the IB, as shown by the density functional theory calculations.[37,38] $\sigma_e$ is the electron-related optical cross-section, which can be obtained as follows in case of deep-level impurities in Si [39]

$$\sigma_e \propto \frac{1}{E_\sigma\sqrt{2\pi}} \int_0^\infty \frac{(\hbar\omega-E_{Te})^{\frac{3}{2}}}{\hbar\omega^3} exp[-\frac{(\hbar\omega-E_{Te})^2}{2E_\sigma^2}]dE \quad (2)$$

where $\frac{(\hbar\omega-E_{Te})^{3/2}}{\hbar\omega^3}$ describes the single discrete deep-level state excitation ($\hbar\omega$ is the photon energy).[39] This this term convolved with a Gaussian distribution of the deep-level states energies, with a mean binding energy $E_{Te}$ and a standard deviation $E_\sigma$, is used to describe the broadening of the deep-level states into an IB with finite width in Te-hyperdoped Si.[36,37] The spectral fit to $\alpha_e$ is shown in Figure 1(g) (red curve). The fitting parameters $E_{Te} = 144 \pm 4\ meV$ and $E_\sigma = 68 \pm 5\ meV$ are obtained with 90.3% confidence intervals by fitting Equation (3) to the absorption coefficient data. This energy position of Te correlates well with the activation energy of the deep Te-levels and agrees with previous works.[29,38,40] The energy probability distribution of deep-level states determined by the spectral fit is plotted in the inset of Figure 1(g) with the *x*-axis being the energetic distance to the CB. The energy probability distribution of the Te deep-levels suggests that the IB is not merged with the CB.

## 2.2. Device characterization: Electrical Properties

For the temperature-dependent electrical measurements, the Te-hyperdoped Si *p-n* intermediate-band Si *p-n* photodiode is placed inside a helium closed-cycle Janis cryostat with a ZnSe window. **Figure 2**(a) depicts the current-voltage (*I-V*) curves at room-temperature for the Te-hyperdoped-Si *p-n* photodiode under dark and white-light illumination conditions with an optical power density of 20 mW/cm². The forward voltage corresponds to the positive bias applied to the bottom contact (*p*-Si substrate). As shown in Figure 2(a), the Te-hyperdoped intermediate-band Si *p-n* photodiode exhibits a rectifying behavior. The ratio forward/reverse current at $\pm 0.5$ V is found to be 357. Further analysis of the rectifying behavior was performed by fitting the dark *I-V* curve using the single-diode equation as follows:[41]

$$I = I_0 \left[ e^{\frac{q(V-IR_S)}{\eta k_B T}} - 1 \right] + \frac{V-IR_S}{R_{shunt}} \qquad (3)$$

where $I_0$ is the saturation current, $q$ is the electron charge, $\eta$ is the ideality factor, $k_B$ is the Boltzmann constant and $T$ is the temperature; whereas $R_s$ and $R_{shunt}$ are the series and parallel resistances, respectively. A room-temperature ideality factor of 2.1 was found by fitting the experimental data. This rectifying behavior is directly related with the *n*-type character of the Te-hyperdoped Si layer,[28] forming a *p-n* junction between the *p*-type Si substrate and the *n*-type Te-hyperdoped layer. At -50 mV reverse bias voltage, the room-temperature dark current density of 0.2 mA/cm$^2$ is lower than that of the reported Au-hyperdoped Si and Se-hyperdoped Si photodetectors.[18,22] Under white-light illumination, an abrupt increase of the reverse current by more than one order of magnitude is observed, showing the operational principle of a photodetector. An open circuit voltage of about 0.2 V was deduced.

Figure 2(b) shows the dark current as a function of the applied voltage at different temperatures ranging from 20 to 300 K. Both the forward and the reverse current increase with increasing temperature. In the reverse bias region, the dark current increases by a factor of $10^6$ from 20 to 300 K. In order to shed light on the transport mechanism in the Te-hyperdoped Si / *p*-Si junction, we fitted the steep part of the forward *I-V* curves to Equation 3. From the fittings, we extracted the temperature dependence of the ideality factor $\eta(T)$ (Figure 2(c)) and of the saturation current $I_0(T)$ (Figure 2(d)). For the ideality factor, we observe two different behaviors depending on the temperature range of operation. For the high-temperature range (T>160 K), the ideality factor shows an almost temperature-independent value between 2.0 and 2.2, suggesting that the generation/recombination processes in the depletion region could be the dominant transport mechanism.[22] On the other hand, for the low-temperature range (T<140 K), the ideality factor is very sensitive to temperature, increasing from 2.2 to 6.7 as the temperature decreases. This temperature dependence of the ideality factor and its value higher than 2 at temperatures below 140 K are indications of a change in the conduction mechanism, i.e. a temperature-independent tunnelling conduction mechanism [42]. Figure 2(d) represents the temperature dependence of the saturation current. We observe that the saturation current also shows two different behaviors. Both follow the Arrhenius law of the form:

$$I_0 = A \times exp(-E_a/kT) \qquad (4)$$

where $A$ is a constant and $E_a$ is the activation energy of the transport mechanism. It is worth noting that both exponential behaviors have been well-fitted in a wide range of current. From fitting to Equation 4 in the high-temperature range (T>160 K), an activation energy of 0.51 eV was deduced. This value is close to about half the silicon band gap (0.56 eV). This supports the previous interpretation of the transport mechanism that is driven by a recombination process in the depletion region in this temperature range.[43] At low temperature (T<160 K), a smaller activation energy of 43 meV was found. Such a small activation energy together with the temperature behavior of the ideality factor support the existence of a tunnelling current [44]. We believe that the origin of this tunnelling current could be related with a multi-tunnelling capture-emission process through the Te-localized states in the junction region, although further work is in progress to elucidate it exactly.

**2.3. Device characterization: Optical Properties**

*2.3.1. Responsivity*

**Figure 3** shows the temperature-dependent spectral responsivity and the band diagram of the Te-hyperdoped Si *p-n* photodiode. The infrared responsivity is estimated at zero bias (i.e. the photovoltaic mode) to prove the pure photovoltaic effect of the photodetector. The Te-hyperdoped Si *p-n* photodiode shows a strong sub-bandgap responsivity up to 5 µm, whereas the responsivity of a commercial Si photodetector reaches the noise floor as expected at wavelengths longer than 1.2 µm. As shown in Figure 3(a), the Te-hyperdoped Si *p-n* photodiode exhibits a room-temperature below-bandgap responsivity of 79 mAW$^{-1}$ at 1.12 µm. Moreover, at the 1.55 µm-telecommunication wavelength, a room-temperature responsivity of around 0.3 mAW$^{-1}$ was obtained, which is comparable to that reported for hyperdoped Si-based photodiodes and solar cells at this wavelength.[45,46] The room-temperature EQE at 1.55 µm was found to be 6×10$^{-4}$, which is comparable to other deep level impurity-hyperdoped Si *p-n* photodiodes.[18,22]

The responsivity of the Te-hyperdoped Si photodetector at different temperatures is displayed in Figure 3(a)-(g). Different behaviors in terms of line shape and responsivity can be identified for three temperature ranges. In the region of 300-200 K, the photoresponse extends up to the MWIR range with a kink in the spectrum, where the responsivity reaches the noise floor at around 4 µm. Interestingly, as the temperature decreases, an additional broad photoresponse band spanning from 1.9 to 3.7 µm is clearly observed (≤ 160 K). Particularly, the responsivity extends well up to 5 µm in the temperature range of 60-20 K. As shown in Figure 3(h) and discussed in the previous work,[34] the sub-bandgap photoresponse observed here corresponds to the excitation of charge carriers from the VB to the CB through the IB (VB→IB (process II) and IB→CB (process III)). Moreover, this observation of sub-bandgap optical transitions indicates that the IB is separated from the CB.

*2.3.2. Detectivity*

The mostly quoted performance merit for a photodetector is its sensitivity, i.e. the input-output signal efficiency compared to the output noise signal. This merit is exemplified as the noise equivalent power (*NEP*), which is defined as the signal power at which the photocurrent can no longer be differentiated from the noise floor, expressed as

$$NEP = \frac{\overline{I_n^2}^{1/2}}{R_{ph}} \qquad (5)$$

Where $R_{ph}$ is the spectral responsivity and $\overline{I_n^2}^{1/2}$ is the root mean square of the total noise current.[8] Here, the major noise sources for the Te-hyperdoped Si photodiode are thermal noise and shot noise. The shot noise is determined by the dark current of the device, $I_{shot} = \sqrt{2qI_{dark}\Delta f}$, where $q$ is the electron charge. The Johnson noise in photodetector at zero bias operation is given by $I_j = \sqrt{\frac{4k_BT}{r_0}\Delta f}$, where $k_B$ is the Boltzmann constant, $T$ is the operation temperature, $\Delta f$ is the electrical bandwidth (1 Hz) and $r_0$ is the resistance product for this detector. Therefore, the specific detectivity ($D^*$) of the photodetector can be expressed as

$$D^* = \frac{\sqrt{A\Delta f}}{NEP} = R_{ph}\sqrt{A}\left[2qI_{dark} + \frac{4k_BT}{r_0}\right]^{-\frac{1}{2}} \qquad (6)$$

where $R_{ph}$ is the responsivity with the units of A/W and $A$ is the photosensitive area (0.082 cm$^2$) of the detector. The calculated specific detectivity ($D^*$) of the device is shown in **Figure 4**. The maximum room-temperature $D^*$ of around $3.2 \times 10^{12}$ cmHz$^{1/2}$W$^{-1}$ is achieved at 1 μm (see Figure 4(a)), which is above two times larger than that of Ag-hyperdoped Si photodetectors.[21] Importantly, the achieved room-temperature $D^*$ at 1 μm is comparable to that of a commercial Si Photodiode (i.e. FDS1010).

Figure 4(b) shows the $D^*$ of the Te-hyperdoped Si photodetector at different temperatures, from 20 to 300 K. The $D^*$ related to the room-temperature sub-bandgap photoresponse is in the range of 10$^{11}$-10$^6$ cmHz$^{1/2}$W$^{-1}$, which is 4 orders of magnitude larger than that of Ti-supersaturated Si photodetector.[19] With increasing temperature, the detector exhibits a decreasing $D^*$, which is mainly due to the increase of the noise as the temperature increases. Figure 4(c) and (d) display the comparison of the spectral $D^*$ of the Te-hyperdoped Si photodetector with a commercial photodetector at 300 K and 77 K, respectively. Notably, at the wavelength range of 1.0-1.9 μm, the room-temperature $D^*$ of this prototype photodetector based on Te-hyperdoped Si exceeds that of a commercial PbSe detector and it is only one order of magnitude smaller than that of a commercial Ge detector. Figure 4(e) displays $D^*$ of the Te-hyperdoped Si photodetector at 1.55 μm and 3.2 μm as a function of temperature. The room-temperature $D^*$ at 1.55 μm and 3.2 μm is around two orders of magnitude lower than that of a commercial Ge photodiode (i.e. FDG03) and an InGaAs photodiode (i.e. FD10D), respectively. However, this performance can be improved by optimizing the manufacturing process. In the future, efforts must be focused towards an advanced-device design to boost the device efficiency. The device architecture can be optimized by proper surface passivation, better optical active area design for improving absorption, different contact geometries and metal electrodes for improving carrier collection, as well as antireflection-coating for improving the responsivity.

Frequency-dependent responsivity is a key factor related to the response time of the devices. Figure 4(f) illustrates the frequency response of the Te-hyperdoped Si photodetector (left axis: responsivity; right axis: $D^*$). The measurement was performed under 1.55 μm-LED illumination with a power density of 1.6 mW/cm$^2$ at different temperatures. The responsivity and $D^*$ have a frequency-dependent response ranging from 1 to 100 kHz. Moreover, the Te-hyperdoped Si photodetector exhibits a cut-off frequency $f_c$ of around 4.2 kHz (this corresponds to a time constant $\tau$ of 37 μs) at room temperature, which is more than two times larger than that of Ti-supersaturated Si photodetectors.[19] This indicates a high bandwidth of the Te-hyperdoped Si photodetector, which also excludes the thermal (bolometric) contribution to the observed responsivity.[8]

*2.3.3. Time-resolved photoresponse*

Another crucial parameter of a photodetector is the transient photocurrent response speed. The response speed of the Te-hyperdoped Si photodetector was measured in the photovoltaic mode at various temperatures by exposing it to a chopped short pulse in the range of 1 Hz to 10 kHz with different 1.55 μm-light intensities. The transient photocurrent was recorded by a digital oscilloscope. As shown in **Figure 5**(a)-(c), the time-dependent photocurrent under varying incident light power with modes of on-off operation at 1 Hz, 1 kHz, and 2 kHz, respectively. In general, a strong and reproducible switching behavior can be observed. Moreover, the variation of the photocurrent with different incident light powers shows an upward trend with the increase of the incident photon energy. The time-dependent photoresponse measured under $P = 0.11$ mW and $f = 1$ kHz at different temperatures is displayed in Figure 5(d). The reproducible on-off behavior is revealed irrespective of temperature, which demonstrates a stable operation regime of the Te-hyperdoped Si photodetector under chopped light illumination. The Te dopants effectively introduce deep-level states inside the Si bandgap and these states act as recombination centers. Moreover, these Te deep-level states are responsible for the rise behavior of the photocurrent. At zero voltage, the room-temperature response time for the rise process was measured to be 39 μs, the time required by the photodetector to generate a photocurrent from 10% to 90% of the peak output value after the light illumination is turned on. The decay time was found to be 42 μs.

## 3. Conclusion

In conclusion, we have fabricated a MWIR infrared photodetector based on Te-hyperdoped Si. We have demonstrated that the Te implanted Si layer is fully recrystallized after PLM and exhibits a Te-mediated broad infrared absorptance up to the MWIR region. In addition, we have investigated the figure of merit of the Te-hyperdoped Si *p-n* photodiodes. From the dark-current measurements, we have observed a transport mechanism based on generation and recombination dominating from 180 to 300 K. The photodiodes have been demonstrated to exhibit a sub-bandgap photoresponse spanning from 1 to 5 μm in a broad temperature range (20-300 K), where a maximum $D^*$ of around $3.2 \times 10^{12}$ cmHz$^{1/2}$W$^{-1}$ has been achieved. At room temperature and under 1.55 μm illumination, the detectors have exhibited a broad bandwidth with a cut-off frequency of 4.2 kHz corresponding to a time constant of 37 μs. The room-temperature rise time and fall time have been determined to be 39 µs and 42 µs, respectively. Te-hyperdoped Si is a promising material for room-temperature Si-based infrared photodetectors. Moreover, the full process, including ion implantation and short-time annealing, is CMOS-compatible. To be competitive with commercial infrared photodetectors, further efforts must be made toward an advanced device design to boost the device efficiency of this prototype of infrared photodetector.

## 4. Experimental Section/Methods

*Material preparation and characterization*: A *p*-type double-side polished Si (100) wafer with a thickness of 380 ± 5 µm and a resistivity of 1-10 Ωcm was implanted with Te ions at room temperature. To obtain a homogeneous distribution of Te ions in the implanted layer, the sample was implanted by two sequential implantations with a dose of $1.6 \times 10^{15}$ cm$^{-2}$ and $6.2 \times 10^{14}$ cm$^{-2}$ and implantation energies of 150 keV and 50 keV, respectively. The implanted layer is 120 nm thick with a peak Te concentration of $2.5 \times 10^{20}$ cm$^{-3}$ (0.5 %), which has been calculated by the SRIM code and then verified by RBS.[47] After ion implantation, the sample was annealed by a spatially homogenized XeCl excimer laser with 308 nm wavelength and 28 ns duration to achieve the re-crystallization of the implanted layer.[32-34] Micro-Raman spectroscopy and TEM were carried out to investigate the structural properties of the doped layers. The Raman spectra were recorded in a backscattering geometry in the wavenumber range of 200 cm$^{-1}$ to 600 cm$^{-1}$ with a continuous 532 nm Nd:YAG laser excitation and detected by a liquid-nitrogen-cooled charge coupled device camera. High-resolution TEM (HRTEM) imaging was performed with an image-C$_s$-corrected Titan 80-300 microscope (FEI). High-angle annular dark-field scanning transmission electron microscopy (HAADF-STEM) imaging and spectrum imaging based on energy-dispersive X-ray spectroscopy (EDXS) were done with a Talos F200X microscope (FEI). Fourier-transform infrared (FTIR) spectroscopy (Bruker Vertex 80v FT-IR spectrometer) was applied to quantify the sub-bandgap absorptance of the obtained samples in the whole MIR (0.05-0.85 eV, λ = 1.4-25 µm). The absorptance (A = 1−T−R) was determined by recording the transmittance and reflectance spectra. Further details about sample preparation and measurements can be found elsewhere. [32-34]

*Device Fabrication*: The structure of the Te-hyperdoped Si *p-n* photodiode comprises a thin Te-hyperdoped Si layer on top of the *p*-Si substrate (as shown in Figure 1 and Figure S1). In detail, the Te-hyperdoped Si sample was immersed into 10% hydrogen fluoride (HF) solution to remove the native-SiO$_2$ layer. Next, a photolithography process was performed to prepare the top electrode. A 0.084 cm$^2$ illuminated area was obtained by defining fingers with a separation of 5 µm resulting in frame-like Au top electrodes on top of the *n*-type Te-hyperdoped Si layer.[33,34] The bottom electrode was made by an In/Ga eutectic layer to form an ohmic contact at the bottom surface with a certain distance from the sample edges in order to reduce possible parasitic electrical conduction through the edges of the sample.

*Device Measurement*: A vacuum pump is used to avoid moisture condensation at low temperatures. A Globar (SiC) source coupled with a TMc300 Bentham monochromator equipped with gratings in Czerny-Turner reflection configuration was used as the infrared monochromatic source. Its intensity is spatially homogenized and was calibrated with a Bentham pyrometric detector. The infrared light emitted from the Globar (SiC) source is modulated by a mechanical chopper at 87 Hz before entering the monochromator. The short-circuit photocurrent was extracted with the help of a SR830 digital signal processing (DSP) lock-in amplifier. For dynamic photoresponsivity experiments such as time-resolved photoresponsivity measurements, a 1.55 µm-light emitting diode (LED) (Thorlabs, 1550E) together with a long-pass filter at 1.3 µm was used. The LED was powered using a homemade current-drivers circuit coupled to the frequency of the output Transistor-Transistor Logic (TTL) signal of the lock-in amplifier. Therefore, the frequency of the TTL signal is adjustable to generate pulsed light and thus perform a frequency scan of the responsivity. Moreover, the output power of the LED is also adjustable with a maximum incident power density of 1.83 mW/cm$^2$. The cut-off frequency of the LED is 0.1 GHz, which is sufficiently higher than that of the Te-hyperdoped Si *p-n* photodetector under investigation. In the time-resolved experiments, a low-noise current amplifier was employed to amplify the photocurrent signal from the device under 1.55 µm-LED excitation. A digital oscilloscope was used to record the time-resolved photocurrent.

**Supporting Information**
Supporting Information is available from the Wiley Online Library or from the author.

**Acknowledgements**

Authors acknowledge the ion implantation group at HZDR (Ion Beam Center) for performing the Te implantations and Romy Aniol for TEM specimen preparation. Furthermore, the funding of TEM Talos by the German Federal Ministry of Education of Research (BMBF; grant No. 03SF0451) in the frame-work of HEMCP is acknowledged. M.W. thanks financial support by Chinese Scholarship Council (File No. 201506240060) and Deutsche Forschungsgemeinschaft (WA 4804/1-1). E.G.H thanks financial support by the Project MADRID-PV (Grant No. MADRID-PV (P2018/EMT-4308)) funded by the Comunidad de Madrid, and by the Spanish MINECO (Ministerio de Economía y Competitividad) under grant TEC 2017-84378-R.

Received: ((will be filled in by the editorial staff))
Revised: ((will be filled in by the editorial staff))
Published online: ((will be filled in by the editorial staff))

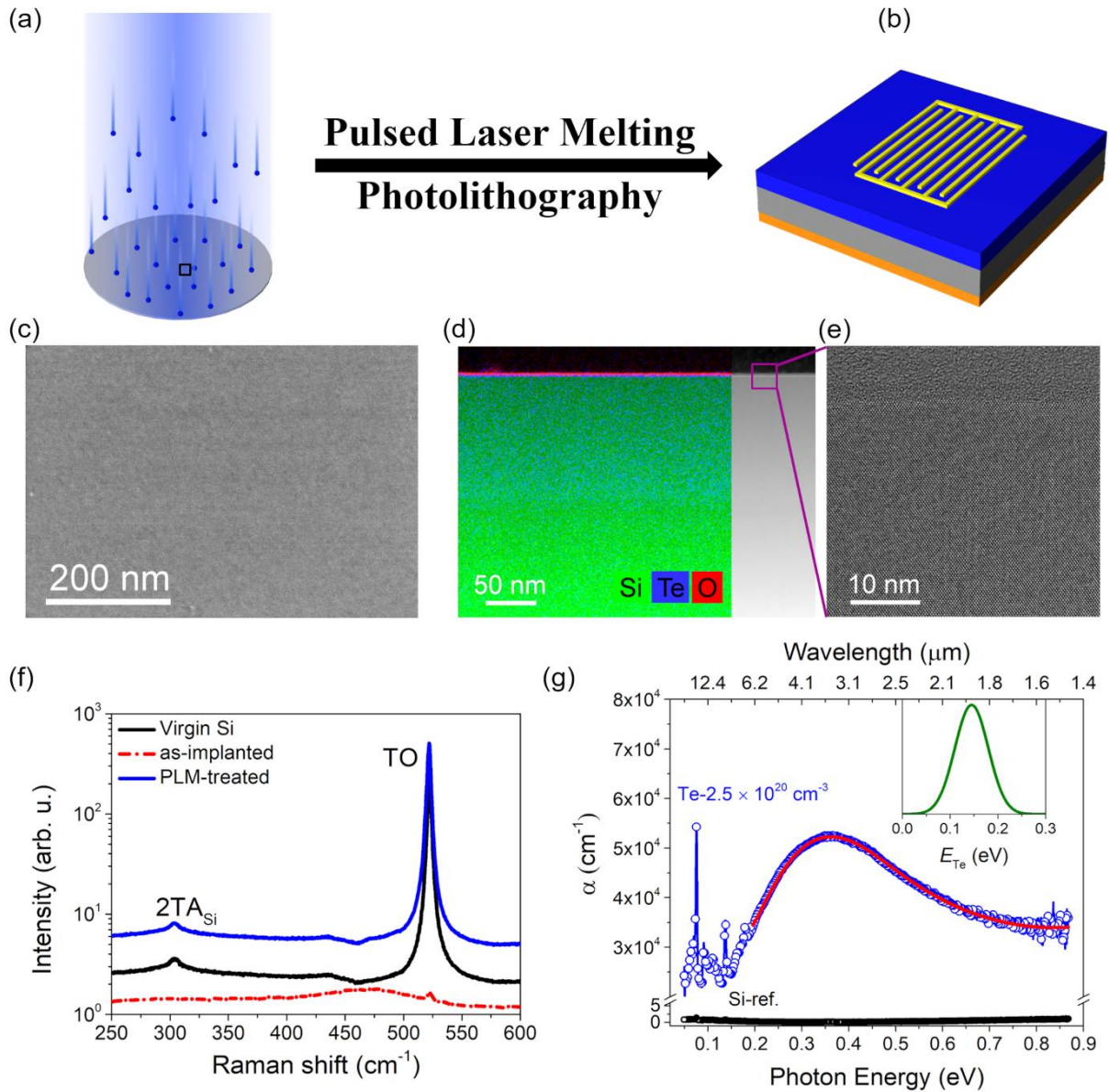

**Figure 1.** The illustration of the Te-hyperdoped Si, including the process flow and the material's properties investigation. (a) Schematic illustration of ion implantation. (b) Schematic illustration of the Te-hyperdoped Si photodetector. (c) Top-view SEM image of the PLM-treated Te-hyperdoped Si layer. (d) Cross-sectional HAADF-STEM image superimposed with the corresponding EDXS element maps (blue: tellurium, green: silicon, red: oxygen); (e) representative HRTEM image for a field of view as depicted by the magenta square in image part (d). (f) Room-temperature micro-Raman spectra of the as-implanted sample and the pulsed laser melting (PLM)-treated Te-hyperdoped Si layer with a Te concentration of 0.50%. Virgin Si is also included for comparison. The spectra have been vertically offset for clarity. (g) Material properties. Optical sub-bandgap absorption spectra from FTIR measurements for virgin Si (black), and the PLM-treated Te-hyperdoped Si with a Te concentration of 0.50% (blue). The inset shows the energy probability distribution of Te deep-level states which is determined by fitting the absorption spectrum with Equation (3) (red). The raw data for the absorptance of sample Te-0.5% can also be found in ref. [43].

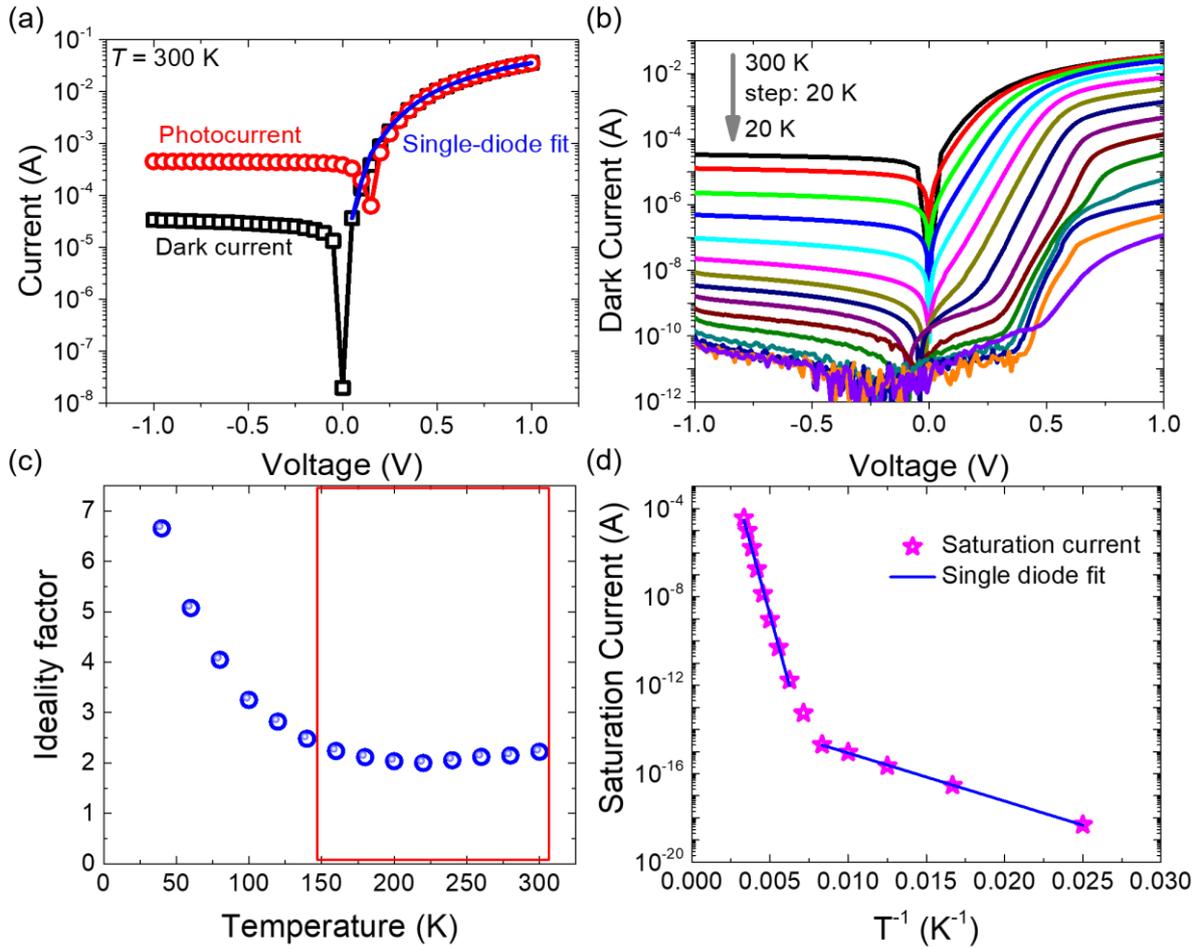

**Figure 2.** (a) Room-temperature dark- and illumination-current-voltage curves (*I-V*) of the Te-hyperdoped Si *p-n* photodiode. The solid line is the single-diode model fitting using Equation (4) (b) Temperature-dependent I-V curves under dark conditions for the Te-hyperdoped Si *p-n* photodiode. (c) Temperature-dependent ideality factor obtained from the single diode model fit. (d) Temperature-dependent saturation current obtained from the fit to equation (4).

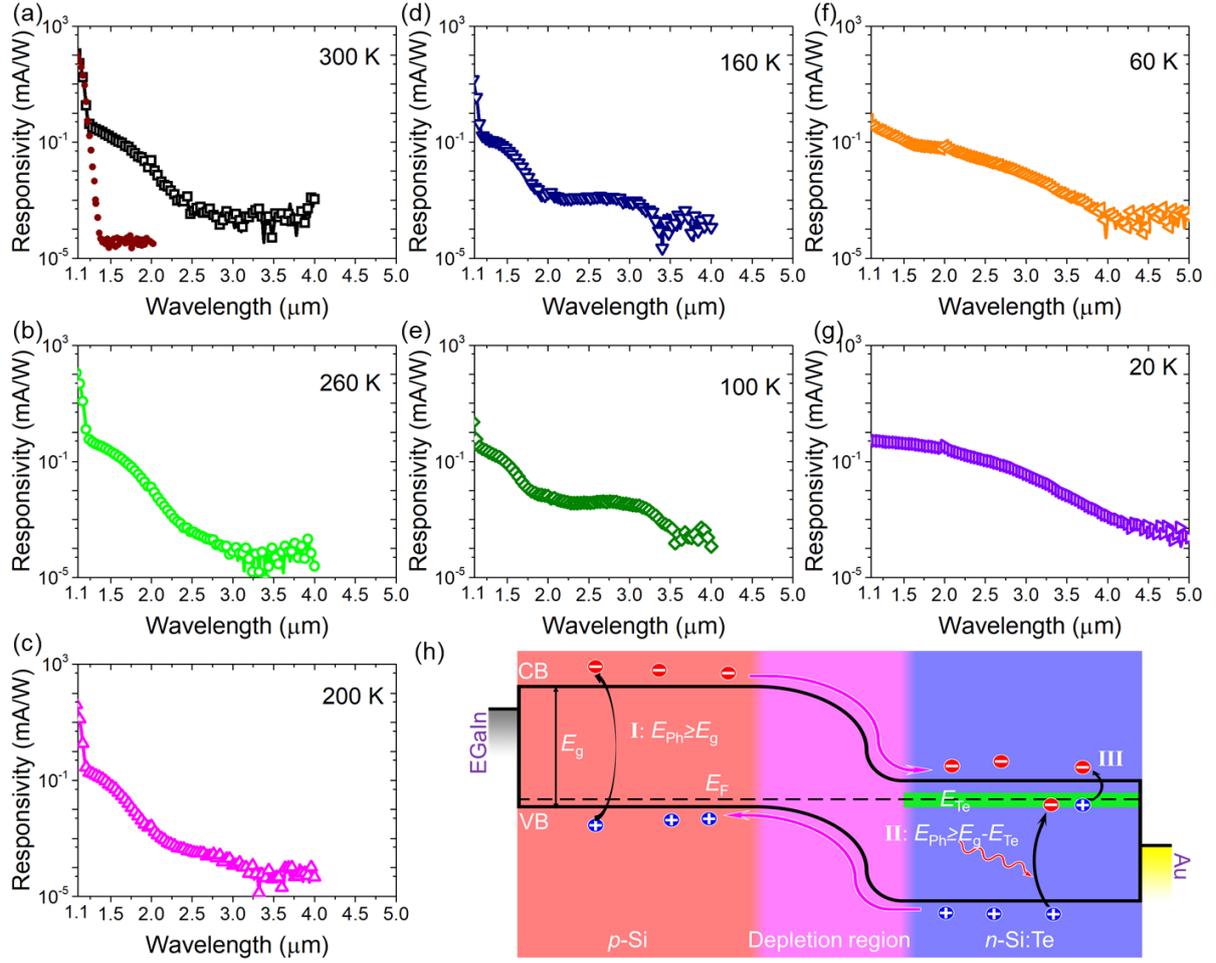

**Figure 3.** (a)-(g) The spectral responsivity measured at zero bias (i.e. photovoltaic mode) for the Te-hyperdoped Si photodetector at different temperatures. The room-temperature spectral responsivity of a commercial Si-PIN photodiode (model: BPW34) is included as a reference (brown short dot). (h) Illustration of the below-bandgap photoresponse in the Te-hyperdoped Si photodetector. Te dopants introduce deep-level states (intermediate band) inside the Si band gap, which facilitate the absorption of photons with sub-bandgap energies. Process I: VB to CB ($E_{ph} \geqslant E_g$); Process II: VB to IB ($E_{ph} \geqslant E_g - E_{Te}$); Process III: IB to CB ($E_{ph} \geqslant E_{Te}$, only measurable at low temperatures where the thermal contribution is neglected).

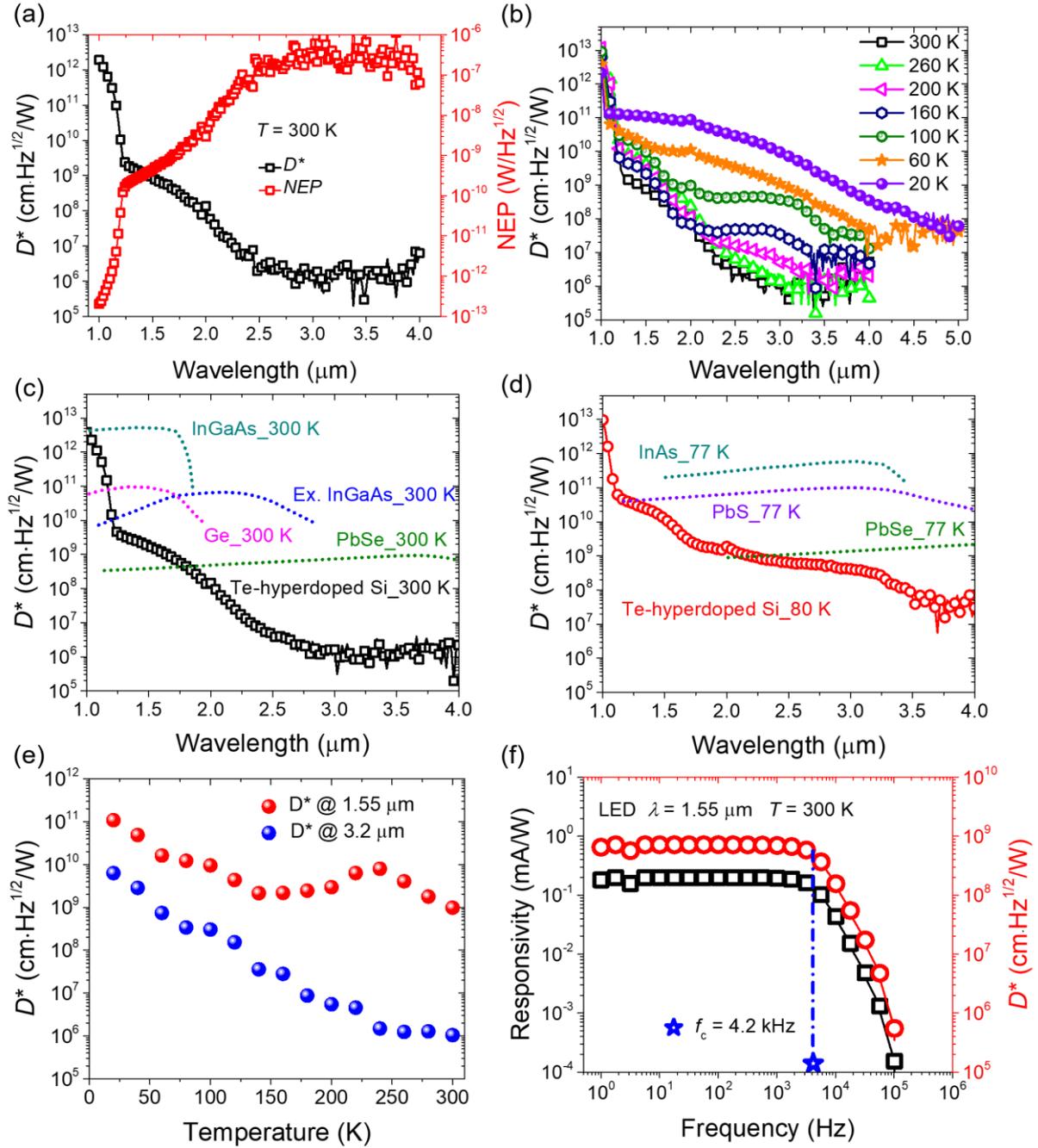

**Figure 4.** (a) Room-temperature wavelength-dependent specific detectivity ($D^*$) and noise equivalent power (NEP) of the Te-hyperdoped Si detector. (b) $D^*$ spectra as a function of wavelength under zero bias at different temperatures from 20 to 300 K. (e) $D^*$ at 1.55 μm and 3.2 μm as a function of temperature. (f) The frequency-dependent photoresponse of the Te-hyperdoped Si photodetector under LED light illumination at a wavelength of 1.55 μm and a power density of 1.6 mW/cm$^2$ at room temperature (left axis: frequency-dependent $R_{ph}$; right axis: frequency-dependent $D^*$.

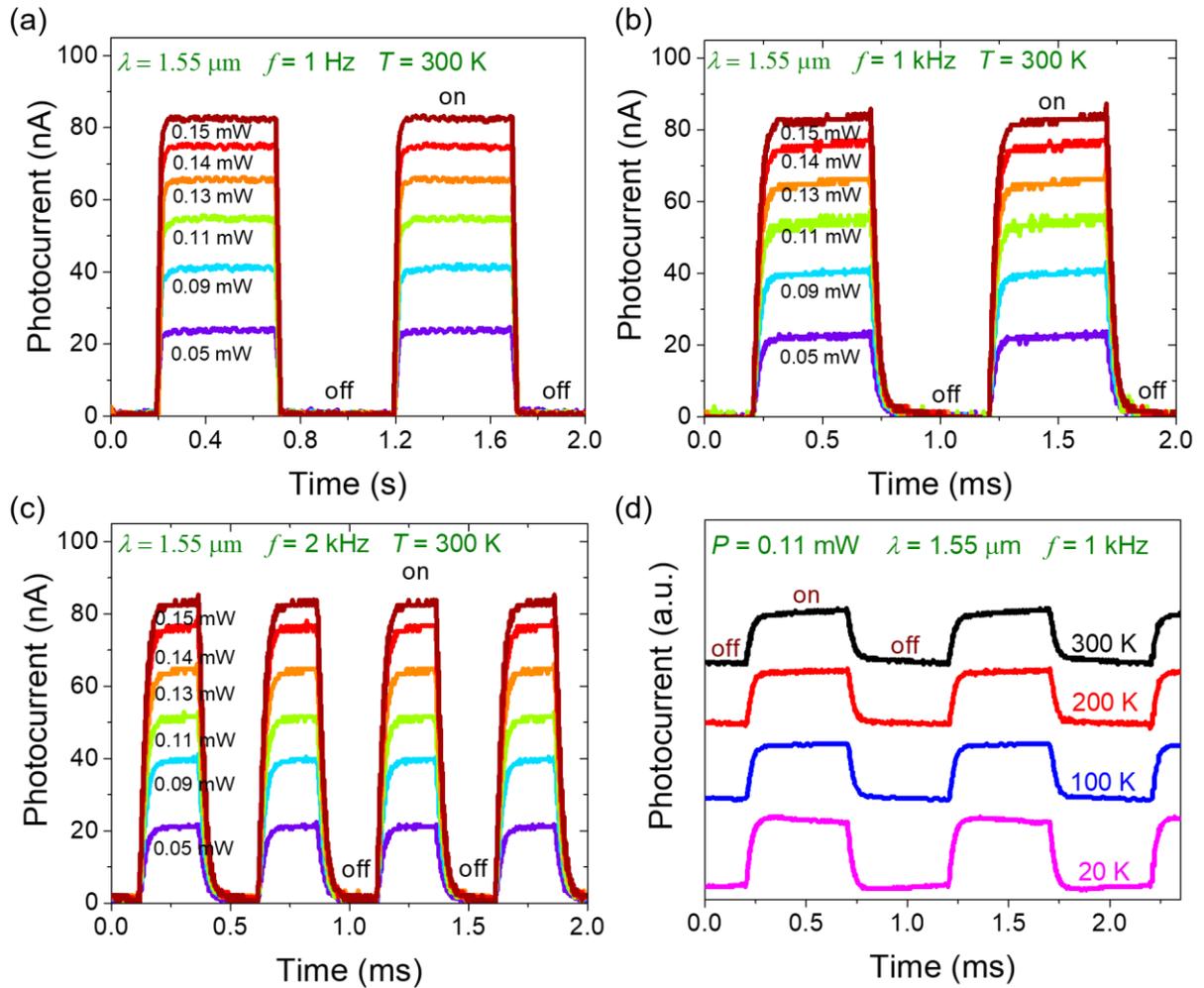

**Figure 5.** Photoresponse of the Te-hyperdoped Si photodetector under the excitation of 1.55 µm LED light illumination measured at zero bias voltage. (a)-(c) The transient photocurrent under varying incident light powers with modes of on-off operation at different frequencies. (d) The transient photocurrent measured at different temperatures with $P$ = 0.11 mW and $f$ = 1 kHz. The curves have been vertically offset for clarity.